\documentclass[twocolumn]{aastex631}

\usepackage{enumerate} 

\usepackage{import}
\usepackage{amsmath}
\usepackage{amssymb}
\usepackage{soul}

\received{January 1, 2999}
\revised{January 2, 2999}
\accepted{January3, 2999}
\submitjournal{ApJ}

\shorttitle{Wavelength Dependence of the S\'ersic Index}
\shortauthors{Martorano et al.}

\begin{document}

\title{Rest-Frame Near-Infrared Radial Light Profiles up to $z=3$ from JWST/NIRCam: Wavelength Dependence of the S\'ersic Index}

\author[0000-0003-2373-0404]{Marco Martorano}
\affiliation{Sterrenkundig Observatorium, Universiteit Gent, Krijgslaan 281 S9, 9000 Gent, Belgium}
\author[0000-0002-5027-0135]{Arjen van der Wel}
\affiliation{Sterrenkundig Observatorium, Universiteit Gent, Krijgslaan 281 S9, 9000 Gent, Belgium}
\author[0000-0002-5564-9873]{Eric F. Bell}
\affiliation{Department of Astronomy, University of Michigan, 1085 South University Avenue, Ann Arbor, MI 48109-1107, USA}
\author[0000-0002-8871-3026]{Marijn Franx}
\affiliation{Leiden Observatory, Leiden University, P.O. Box 9513, 2300 RA, Leiden, The Netherlands}
\author[0000-0001-7160-3632]{Katherine E. Whitaker}
\affiliation{Department of Astronomy, University of Massachusetts, Amherst, MA 01003, USA}
\affiliation{Cosmic Dawn Center (DAWN), Niels Bohr Institute, University of Copenhagen, Jagtvej 128, Kø benhavn N, DK-2200, Denmark}
\author[0000-0001-6843-409X]{Angelos Nersesian}
\affiliation{Sterrenkundig Observatorium, Universiteit Gent, Krijgslaan 281 S9, 9000 Gent, Belgium}
\author[0000-0002-0108-4176]{Sedona H. Price}
\affiliation{Department of Physics and Astronomy and PITT PACC, University of Pittsburgh, Pittsburgh, PA 15260, USA}
\author[0000-0002-3930-2757]{Maarten Baes}
\affiliation{Sterrenkundig Observatorium, Universiteit Gent, Krijgslaan 281 S9, 9000 Gent, Belgium}
\author[0000-0002-1714-1905]{Katherine A. Suess}
\affiliation{Department of Astronomy and Astrophysics, University of California, Santa Cruz, 1156 High Street, Santa Cruz, CA 95064 USA}
\affiliation{Kavli Institute for Particle Astrophysics and Cosmology and Department of Physics, Stanford University, Stanford, CA 94305, USA}
\author[0000-0002-7524-374X]{Erica J. Nelson}
\affiliation{Department for Astrophysical and Planetary Science, University of Colorado, Boulder, CO
80309, USA}
\author[0000-0001-8367-6265]{Tim B. Miller}
\affiliation{Department of Astronomy, Yale University, 52 Hillhouse Avenue, New Haven, CT 06511, USA}
\author[0000-0001-5063-8254]{Rachel Bezanson}
\affiliation{Department of Physics and Astronomy and PITT PACC, University of Pittsburgh, Pittsburgh, PA 15260, USA}
\author[0000-0003-2680-005X]{Gabriel Brammer}
\affiliation{Cosmic Dawn Center (DAWN), Niels Bohr Institute, University of Copenhagen, Jagtvej 128, Kø benhavn N, DK-2200, Denmark}

\correspondingauthor{Marco Martorano}
\email{marco.martorano@ugent.be}
\begin{abstract}
    We examine the wavelength dependence of radial light profiles based on S\'ersic index $n$~measurements of 1067 galaxies with M$_*\geq$ 10$^{9.5}$M$_\odot$ and in the redshift range $0.5<z<3$.  The sample and rest-frame optical light profiles are drawn from CANDELS$+$3D-HST; rest-frame near-infrared light profiles are inferred from CEERS JWST/NIRCam imaging. $n$ shows only weak dependence on wavelength, regardless of redshift, galaxy mass and type: on average, star-forming galaxies have $n=1-1.5$~and quiescent galaxies have $n=3-4$ in the rest-frame optical and near-infrared. The strong correlation at all wavelengths between $n$ and star-formation activity implies a physical connection between the radial stellar mass profile and star-formation activity.  The main caveat is that the current sample is too small to discern trends for the most massive galaxies (M$_*>10^{11}~M_\odot$~).

\end{abstract}


\section{Introduction}

    James Webb Space Telescope (JWST) NIRCam imaging is providing us for the first time with spatially resolved, rest-frame near-infrared structural information of galaxies at high redshift (up to z$\sim$3) with an angular resolution unachievable from ground-based observatories. Previous, extensive Hubble Space Telescope (HST) surveys have produced a detailed picture of the rest-frame UV and optical structural properties and their correlations with other galaxy properties such as stellar mass and star-formation rate \citep[i.e.,][]{van-der-wel14, van-der-wel14a, lang14, shibuya15, nelson16}, but only at longer wavelengths is the stellar body revealed without being significantly hampered by the effects of stellar population and attenuation variations across galaxies.
    
    The broad wavelength coverage together with the high angular resolution offered by the synergy between HST  and JWST observations in the near-IR allows one to examine the wavelength dependence of the S\'ersic index ($n$). For the first time we can determine for galaxies at large look-back time whether any trends in the rest-frame near-IR with stellar mass and star-formation activity are substantially different from those seen in the rest-frame optical/UV.

    Both non-parametric and parametric methods have been widely used to examine galaxy structure: see \citet{conselice14} for a comprehensive review on non-parametric methods and \citet{Whitney21} for a recent application to high-redshift galaxies. Among parametric methods, S\'ersic profile fits have become the standard as they are conveniently measured with codes such as \textsc{galfit} \citep{peng02, peng10a}; this is the approach we use in this study.
    
    As stellar populations and dust attenuation determine the light distribution, the measured structural properties are generally observed to depend on wavelength \citep[e.g,][]{de-jong96a,de-jong96b, Kelvin12, haussler13, vika13, Pastrav13, vulcani14, kennedy15, baes20, nersesian23}.
    \cite{Kelvin12} used the low-redshift ($z<0.3$) GAMA  survey \citep{driver09} to examine the dependence of the S\'ersic index on wavelength, splitting the galaxy sample by visual morphology (disc galaxies and spheroidal galaxies). They find a mild dependence on wavelength for spheroidal galaxies and a somewhat stronger correlation for disc galaxies. \citet{Kelvin12} argue that the trend is to be expected, as spiral galaxies will have redder bulges, which tend to have $n>1$. Despite this trend, the difference in S\'ersic index between disc and spheroidal galaxies persists even in the $K_s$ band ($n\lesssim 2$~vs.~$n\approx3.5$).  
    
    Since the morphological class and structure/concentration (as parameterized by S\'ersic index) are intrinsically connected and, from an empirical perspective, derived from the same information, we will instead examine the wavelength dependence of S\'ersic $n$ separating galaxies by star-formation activity, which is estimated independently from the S\'ersic profile. Moreover, thanks to the revolutionary NIRCam imaging we can for the first time extend such an analysis to redshift $z=2$ and beyond. The goal of this study is to measure the evolution of the S\'ersic index with redshift in the rest-frame near-infrared and examine the wavelength dependence of the S\'ersic index across the redshift range $0.5<z<3$. One question of specific interest is whether the difference between the S\'ersic index for quiescent and star-forming galaxies, seen in the rest-frame optical \citep[e.g.,][]{blanton03a, blanton09, bell12}, persists in the rest-frame near-IR. If not, then the star-formation activity itself is the cause of the apparent difference in structure (while the underlying mass profile shows no such difference). Alternatively, if the difference persists, then there is a physical correlation between the shape of the radial stellar mass profile and star-formation activity.
    
    In this work we take advantage of the JWST/NIRCam \citep{rieke05} imaging provided by the CEERS program \citep{Finkelstein23} and model the light profiles of 1067 galaxies in the redshift range $z=0.5-3$ with stellar masses $M_*\geq10^{9.5}$M$_{\odot}$. We also compare our results with previous studies \citep{Kelvin12, van-der-wel12} to exploit strengths and weaknesses of our study.
    
    The paper is structured as follows: in section \ref{sec: Data} we describe the datasets used in this paper, section \ref{sec: Results} contains the results of this work divided into the different dependencies of S\'ersic indices and a comparison with literature results. This will be followed by a discussion in section \ref{sec:  Conclusions-Discussion} and finally, in section \ref{sec: Conclusions}, we will sum up the content of the paper and draw our conclusions.

    We assume a flat $\Lambda$CDM cosmology with H$_0$=70km$\,$s$^{-1}$ Mpc$^{-1}$ and $\Omega_{\rm{m}}$=0.3.

\section{Data and Sample Selection\label{sec: Data}}
    In this section we construct a redshift and stellar mass-selected sample from pre-existing catalogs in the Extended Groth Strip \citep[EGS,][]{davis07}, and derive S\'ersic profile fits from the recently acquired JWST/NIRCam imaging from CEERS \citep{Finkelstein23}.

\subsection{Imaging and Photometry}\label{sec:phot}
    \citet{skelton14} and \citet{whitaker14} presented HST, Spitzer and ground-based imaging data to construct a widely used, homogeneous photometric catalog from the UV to 24$\mu$m for the five extragalactic deep fields targeted by CANDELS \citep{koekemoer11,grogin11} and 3D-HST \citep{brammer12} which \citet{Leja20} used to estimate redshift, stellar masses, star-formation rates and dust attenuation parameters with the \textsc{Prospector} SED fitting code \citep{Johnson17, Johnson21}. In this paper we use the \citet{Leja20} catalog of redshifts, stellar masses and star-formation rates for galaxies at redshift $0.5<z<3$. 

    We first cross-match this catalog with \cite{van-der-wel12} which will be used as a reference for comparisons, and then with sources in the recent JWST/NIRCam data taken as part of the Cosmic Evolution Early Release Science Survey (CEERS) program  \citep{Finkelstein17,Finkelstein23} in the EGS, one of the 5 CANDELS fields. We note (and correct for) a systematic shift in declination of $+0.19"$ between the \citet{skelton14} coordinates and the coordinates in the CEERS mosaics, which we identified with the Python library for Source Extraction and Photometry \textsc{sep} \citep{bertin96, barbary16, barbary18}.
    We identify 2684 out of 63,413 galaxies in the \citet{Leja20} catalog falling within the NIRCam footprint, of which 1216 are above our adopted stellar mass limit of $M_*=10^{9.5}$M$_{\odot}$, which is the stellar mass completeness limit of the photometric catalog at $z\sim3$ \citep{tal14}. 97\% of these objects have signal-to-noise ratio $>$50 in all JWST/NIRCam filters, which is the requirement for unbiased S\'ersic index measurements \citep{van-der-wel12}. This is not the case for the shallower HST/WFC3 data, but those data do not play a major role in this work.

    CEERS provides imaging in seven near-IR filters, specifically F115W, F150W, F200W in the Short Wavelength channel of NIRCam and F277W, F356W, F410M, F444W through the Long Wavelength channel. We use Stage 2b calibrated images already background subtracted available on the MAST archive additionally processed with the \textsc{Grizli} software \citep{Brammer19} to obtain aligned imaging, weight and segmentation mosaics out of the 4 pointings available when we started this work \citep[also see][]{naidu2022}.

    For convenience, the available HST/ACS (F606W and F814W) and HST/WFC3 (F125W, F140W and F160W) data are re-reduced with \textsc{Grizli} to produce mosaics that are aligned with the NIRCam mosaics from CEERS.

\subsection{S\'ersic Profile Fits \label{sec: Sersic Profile Fits}}
    For each of the 1216 target galaxies with NIRCam imaging
    and for each of the seven JWST filters and five HST filters, we create square cutouts with size ten times the effective radius in pixels measured from the F160W CANDELS imaging \citep{van-der-wel12}.
    We set a lower limit of 63 pixels ($\sim$2.5") and an upper limit of 200 pixels for sources with $m_{\rm{F160W}}>22.5$. The upper limit is doubled for sources brighter than this threshold to ensure accurate background estimates. All objects in the cutout that are brighter than or less than 1 magnitude fainter than the target are assigned their own S\'ersic profile and simultaneously fitted. All other sources in the \citet{van-der-wel12} catalog or in the CEERS segmentation map are masked. 
    
    We use \textsc{GalfitM} \citep{haussler13, vika13} to perform the profile fits simultaneously for all JWST/NIRCam filters and simultaneously for all HST/ACS and HST/WFC3 filters. The model S\'ersic profiles are convolved with the publicly available model PSFs\footnote{\url{https://github.com/gbrammer/grizli-psf-library/tree/main/ceers}} drizzled with the same drizzle parameters as those used to create the mosaics. 
    The square root of the inverse of the weight map is used as the noise map in \textsc{GalfitM}.
    The background, as well as the center of the galaxy, the total magnitude, S\'ersic index and effective radius are left as free parameters of the fit, allowing those to vary independently from filter to filter. 
    The axis ratio and position angle are the additional free parameters, but these are constrained to have the same value for all filters in the simultaneous fits. 
    
    We set the following constraints on the fit's parameters:
    \begin{itemize}
        \item $0.2\leq$ S\'ersic index $\leq12$
        \item $0.01\leq$ Re [pix] $\leq150$
        \item x and y coordinates have to be within 5 pixels from the \cite{van-der-wel12} corrected position.
    \end{itemize}

    Following \citet{van-der-wel12} we assign a formal random uncertainty of 0.1 dex on the S\'ersic index for objects with signal-to-noise ratio S/N$=50$, and scale the (linear) uncertainty with (S/N)$^{-1/2}$. The S/N is calculated by summing the image and inverse weight maps across the object segment in the segmentation map.

\begin{figure*}[!t]
        \epsscale{1.1485}
        \plotone{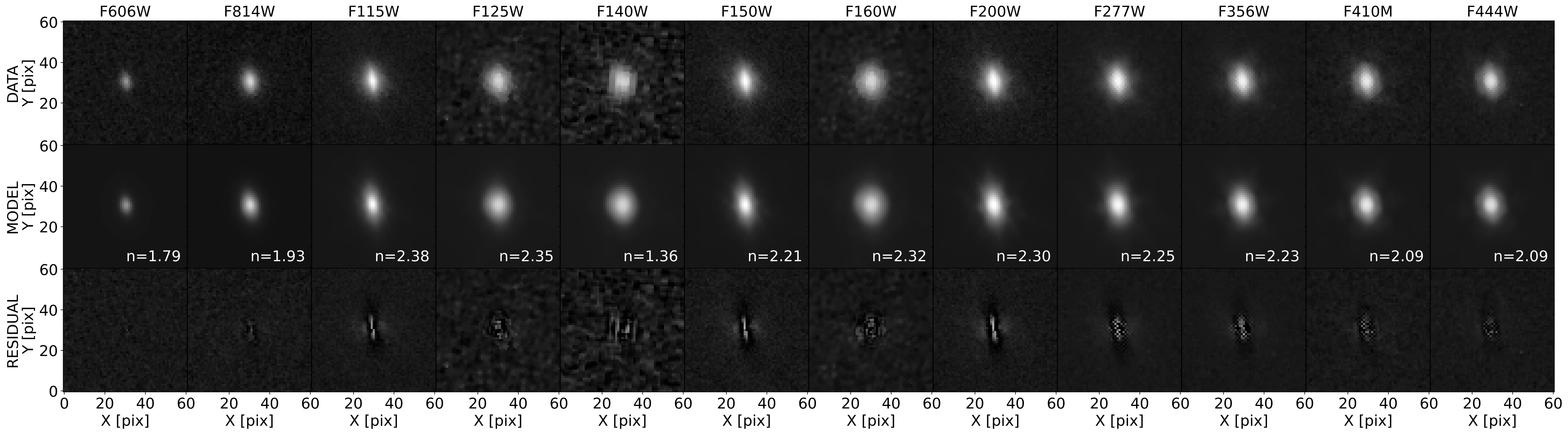}
        \plotone{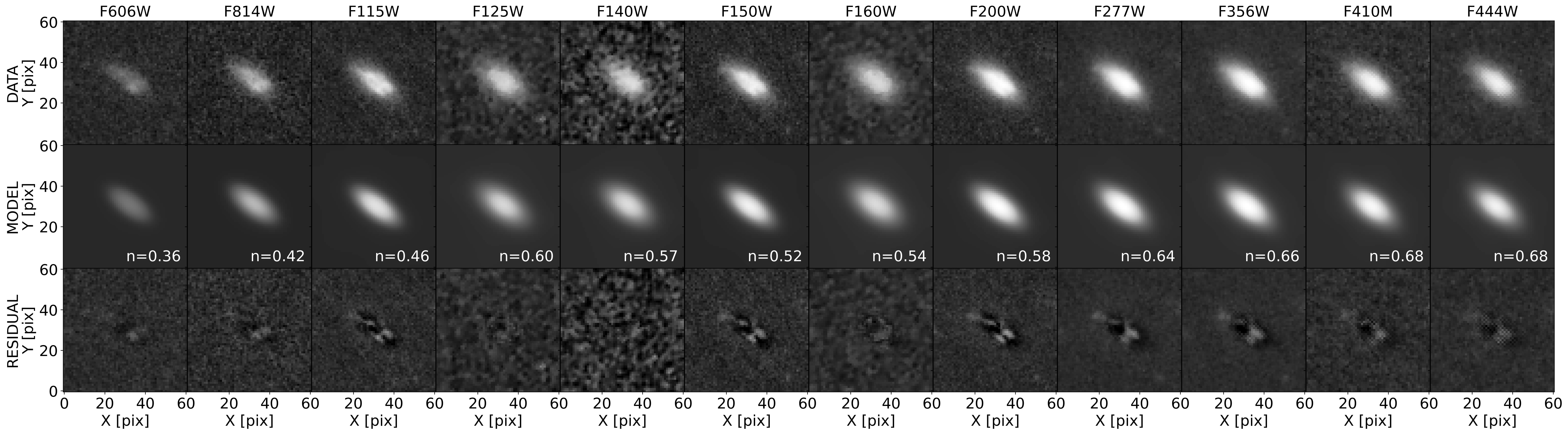}
        \plotone{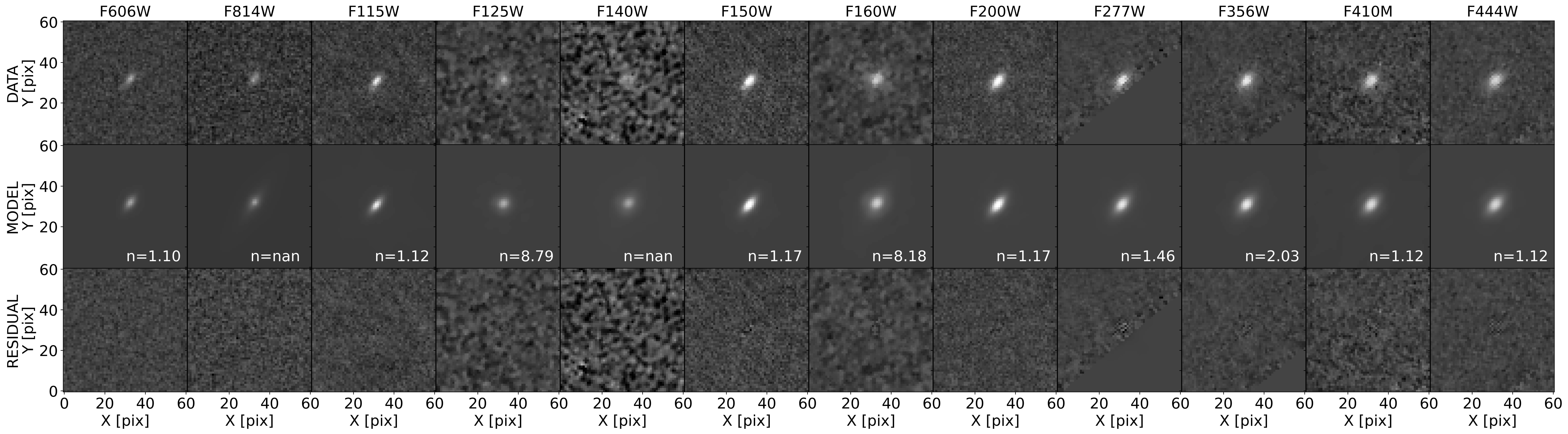}
        \plotone{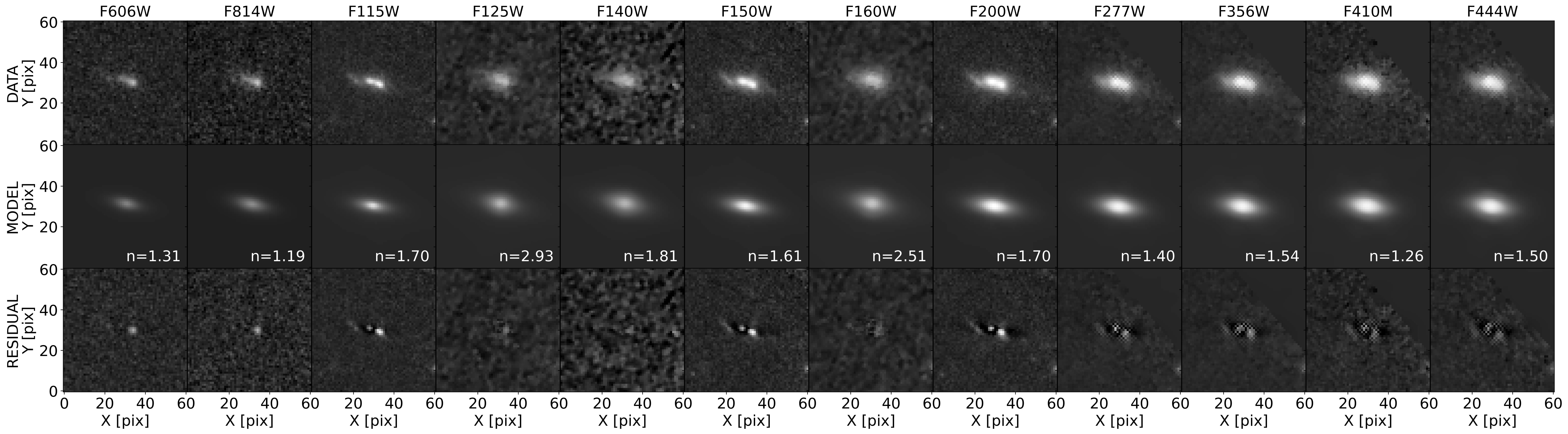}
        \caption{\emph{First panel:} Galaxy 5751 in \cite{Leja20} catalog with M=10$^{9.84}$M$_{\odot}$ and redshift z=0.74, \emph{Second panel:} Galaxy 11438 with M=10$^{9.60}$M$_{\odot}$ and redshift z=0.98, 
        \emph{Third panel:} Galaxy 10662 with M=10$^{9.53}$M$_{\odot}$ and redshift z=2.49, \emph{Fourth panel:} Galaxy 14128 with M=10$^{9.60}$M$_{\odot}$ and redshift z=2.98.
        For each of the four sets, the first row shows the original image, the second row the model used and the third row the residuals. Per each galaxy the color of the panels spans the same scale in all the filters. Cutouts have been cropped with respect to those adopted for the fit to enhance galaxy visibility.
        Together with the model, in the bottom right corner, is shown the measured S\'ersic index in each filter. In case the fit of a filter failed, the value of S\'ersic index reported is \emph{nan}.
        \label{fig:cutouts}
        }
    \end{figure*}

    In Figure \ref{fig:cutouts} we show the data, model and residual of four $10^{9.5}-10^{9.8}M_{\odot}$ galaxies respectively at redshift 0.74, 0.98, 2.49 and 2.98. Even these low-mass galaxies are detected with high significance and are well resolved by JWST/NIRCam filters, even when HST does not.

    In order to calculate the S\'ersic index at specific rest-frame wavelength values we fit for each object a 2nd-order Chebyshev polynomial to the independently measured S\'ersic $n$ values across all filters (but separately for JWST and HST to exploit differences between the instruments) with the uncertainties as weight factors. For each galaxy we fit the polynomial only to those filters where the galaxy was detected (i.e. does not fall in gaps of the NIRCam detectors) and with a converged value of $n$. We reject from our sample galaxies for which less than 3 filters are available to fit the polynomial. Among the JWST fits, just one galaxy does not satisfy this condition, while for the HST fits we reject 35 galaxies.
    
    The adopted uncertainty for S\'ersic indices at specific rest-frame wavelength recovered from the Chebyshev polynomial is that of the filter nearest in pivot wavelength. 
    The differences between the directly measured S\'ersic index values and the polynomial values are small, typically $<5\%$ and within the error bar.

    We also remove from our sample those 113 objects for which the S\'ersic index reached the $n=0.2$ or $n=12$ constraint for either of the JWST/NIRCam or HST/ACS filters. The WFC3 filters overlap in wavelength with the NIRCam/SW channel, so that such a rejection is not needed.

    This leaves us with a final sample of 1067 galaxies in the redshift range $0.5<z<3$ and M$_*\geq10^{9.5}$M$_{\odot}$. The rejected galaxies do not severely bias the sample as they do not occupy a particular region of the parameter space in terms of redshift or stellar mass (in the appendix \ref{Appendix} we show the stellar mass and sSFR distribution of the selected sample and of the rejected galaxies as a function of redshift).

    \begin{figure}[!t]
        \epsscale{1.15}
        \plotone{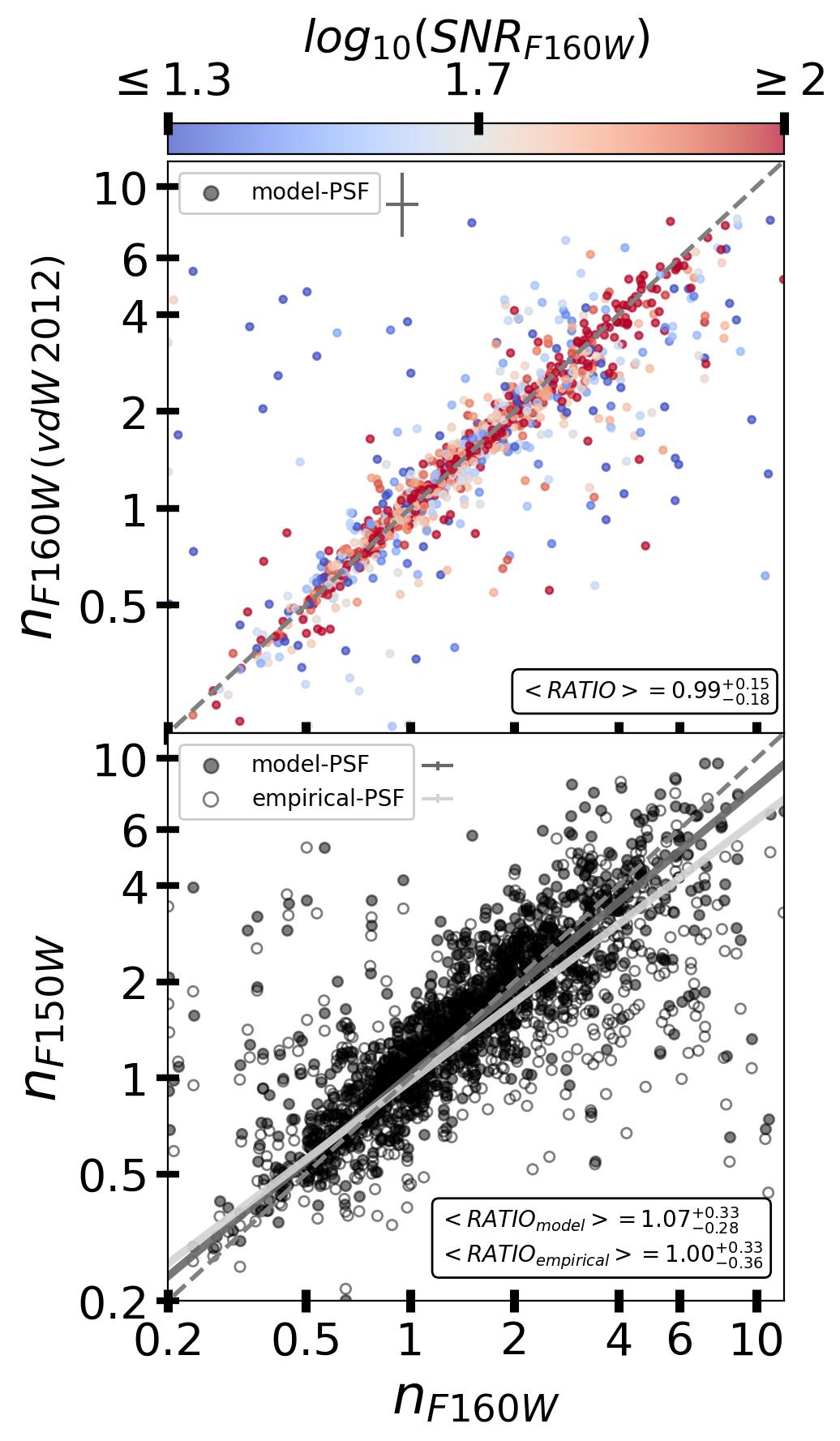}
        \caption{\emph{Upper panel:} Comparison between HST/F160W S\'ersic indices from \cite{van-der-wel12} (y-axis) and HST/F160W S\'ersic indices derived in this paper (x-axis) color-coded by the $log_{10}(SNR)$.  \emph{Lower panel:} Comparison of S\'ersic indices from JWST/NIRCam filter F150W and HST/WFC3 filter F160W (this paper). Filled circles represent F150W estimates with the WebbPSF-based model-psf while open circles show F150W estimates with the empirical psf. Dark-grey and light-grey solid lines show linear fits respectively on the model-psf and empirical psf datasets after 3-sigma clipping. In the lower right corner of each panel are reported the median ratios and the 16-84 percentile intervals. Next to the legends we show the median uncertainties. The first panel shows the agreement with literature results is remarkable, while the second highlights how differences between empirical and model-PSF arise mostly for high values of $n$.
        \label{fig:sersic160_sersic150}
        }
    \end{figure}

    To check whether the size adopted for cutouts is large enough to properly retrieve S\'ersic profiles, we compare our size measurements with those presented in \cite{van-der-wel12} finding no systematic differences.

    To ensure that the background estimate converges for the sizes of the cutouts, we compare the magnitudes with those obtained from cutouts half the size and, for a subset of the 50 most massive galaxies, the magnitudes obtained with cutouts double the size. From this exercise we conclude that the magnitudes are accurate to within 0.05 mag (and S\'ersic indices to within 0.03 dex), indicating the systematic uncertainties in the background estimates do not play a significant role. To check for spatial variations of the background within cutouts we compared the standard deviation of pixels in the residuals with the median value in the noise-maps finding a good agreement between the two. This confirms that background spatial variations due to close bright sources or diffuse halos do not contribute significantly to the uncertainties.
    
    Likewise we test whether constraining the position angle and axis ratio to the same value for all filters leads to systematic errors. With those parameters free to vary from filter to filter we find only small scatter  in S\'ersic $n$.

    To test the reliability of our new HST-based S\'ersic profile fits we compare in the upper panel of Figure \ref{fig:sersic160_sersic150} our measurements of S\'ersic index in the HST/WFC3 filter F160W with the previously published values from \cite{van-der-wel12} inferred from the same data. Due to slight differences in the segmentation of objects and in methodology there is scatter and a fair number of outliers appear ($\approx30\%$ of the sample have $\Delta$log($n$) $>0.08$dex with no preferred properties, where $\Delta$log($n$) is defined as $|log_{10}(n_{F160W(vdw2012)})-log_{10}(n_{F160W})|$), but the systematic offset is negligible and the 1$\sigma$ scatter small ($\sim16\%$). Galaxies with $\Delta$log($n$) $>0.08$dex are characterized by a median S/N$_{F160W}$ that is half of that of galaxies with smaller $\Delta$log($n$), suggesting low S/N might be an important source of scatter in this plot.
    
    In the bottom panel of Figure \ref{fig:sersic160_sersic150} we show a comparison of NIRCam/F150W and HST/F160W S\'ersic indices retrieved in this work. From this figure, we can infer that despite being shallower than JWST/F150W, HST/F160W was already deep enough to properly recover S\'ersic indices for the whole population \citep{van-der-wel12,Nedkova21}. Indeed, quantitatively, $n_{F150W}$ is just $\sim$7\% systematically larger than $n_{F160W}$. However, the large scatter observed reflects how the improved depth of F150W highlights new features that can change the S\'ersic index of some galaxies. An example of such differences can be observed in the last two panels of Figure \ref{fig:cutouts}.

    The scatter is larger by about a factor 2 compared to what one expects based on the formally adopted measurement uncertainties: the total uncertainty (on the ratio of the two independently measured S\'ersic indices shown in the figure) is as much as 50\%. That said, the random uncertainties on the CEERS-based measurements are typically a factor 3-4 smaller than those from CANDELS due to the increased S/N, leading us to conclude that the CEERS-based measurements have a variance that is one-quarter of the total variance seen in Fig.~\ref{fig:sersic160_sersic150}. The error budget is an important issue that needs to be examined in detail, but it is also beyond the scope of this paper. Future improvements in the PSF model and background subtraction techniques are certain to improve the data analysis.

    To test whether the choice of the PSF used to fit NIRCam mosaics affects our results, we repeat the profile fits with an empirical PSF: a star visually identified in the NIRCam mosaics. 
    The star is chosen to be faint (and therefore noisy) in order to avoid the well-known saturation issues in the NIRCam imaging \citep[for an in-depth analysis of the JWST/NIRCam PSF see e.g.,][]{nardiello22, weaver23, zhuang23}. The difference between the model-PSF and the empirical-PSFs is most pronounced in the Short Wavelength channel, where we see a decrease of $\sim 25\%$ in $n$ for objects with $n>3$. Low-$n$ and Long Wavelength channel estimates are not systematically affected. This bias does not introduce large significant uncertainties in the following analysis and our conclusions do not depend on the choice of PSF. 
    Solid dark(light)-grey lines show linear regression fits to the $n$ values inferred with the model (empirical) PSF, using 3$\sigma$ clipping. The empirical PSF results shows systematic offsets at low and high $n$, whereas the model PSF results do not, implying that they suffer less from systematic uncertainties. The results based on the model PSF are used in the remainder of this paper.

\subsection{Local Comparison Sample from GAMA}\label{sec:gama}
    
    A $z\leq0.3$ comparison sample is drawn from GAMA. For the GAMA I dataset we use the \textsc{GalfitM} profile fits obtained from the SDSS \emph{ugriz} imaging and the UKIDSS-LAS YJHK imaging \citep{lawrence07,lawrence12}.\footnote{\url{http://www.gama-survey.org/dr4/schema/table.php?id=578}}

    \textsc{MAGPHYS} \citep{cunha08} stellar mass and star-formation rate estimates from GAMA II \footnote{\url{http://www.gama-survey.org/dr4/schema/table.php?id=545}} are used to construct the comparison sample of $\sim23600$ galaxies with $M_*\geq10^{9.5}$M$_{\odot}$ and redshift $z\leq0.3$. There may be systematic differences between MAGPHYS and \textsc{Prospector}-based parameter estimates, for the purpose of this study these are not significant: we study trends with mass and bulk redshift evolution so that 0.1-0.2 dex differences in $M_*$ and SFR do not matter.

    \begin{figure*}[ht]
        \epsscale{1.15}
        \plotone{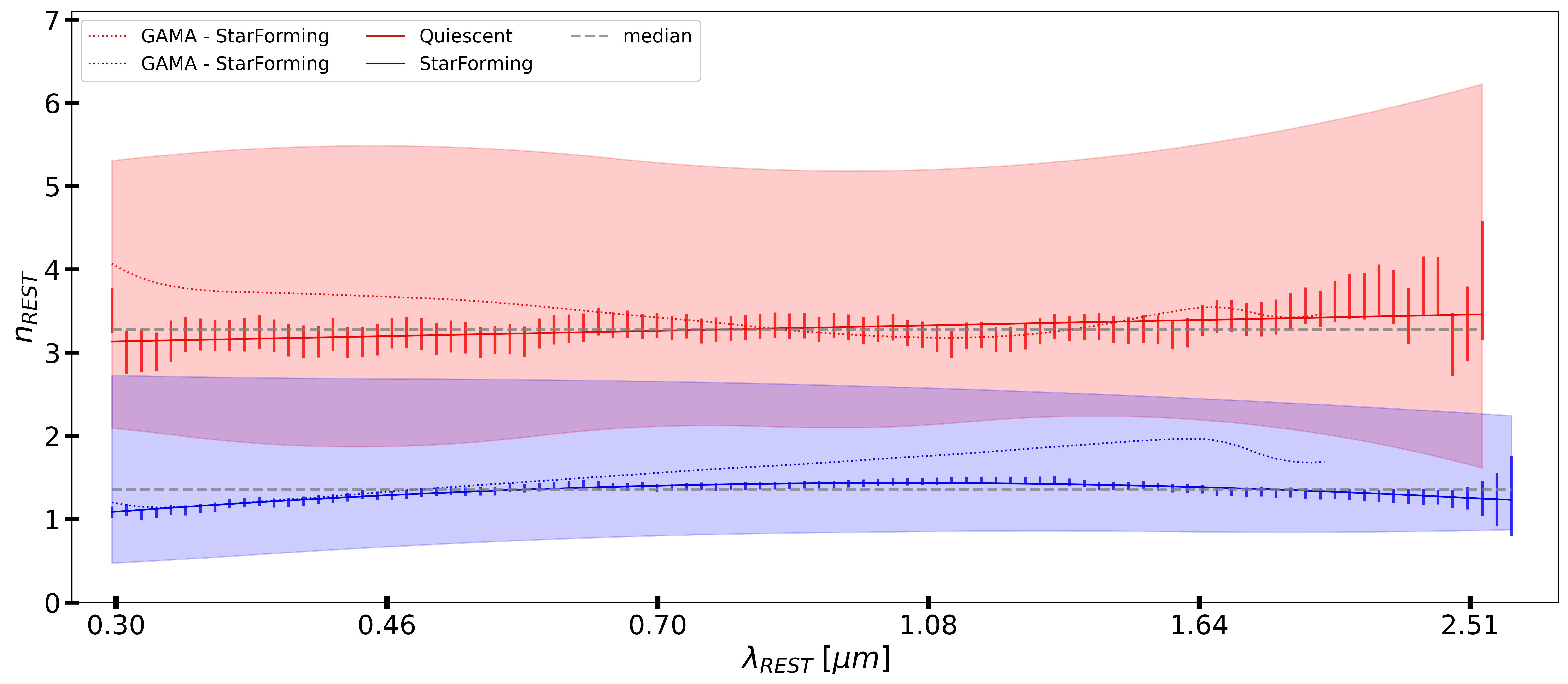}
        \caption{Median S\'ersic index (and 16-84 percentile ranges as shaded areas) of the quiescent and star-forming population as a function of rest-frame wavelength. Errorbars indicate the statistical uncertainty on the median ($\sigma_n/\sqrt{N}$) at log-spaced wavelengths. Solid lines show an spline regression to the medians as outlined in the text. Dashed gray horizontal lines represent the overall median, integrating over wavelength. The dotted lines show the low-$z$~GAMA comparison sample. For both star-forming and quiescent galaxies we see no strong change in $n$ with $\lambda_{REST}$ from the nUV to the nIR.
        \label{fig:sersicVSrestSF}
        }
    \end{figure*}

\section{Results \label{sec: Results}}

    In this section we investigate the wavelength and mass dependence and redshift evolution of the S\'ersic index, based on the 1067 galaxies in the redshift range $0.5<z<3.0$ and with stellar masses $M_*\geq10^{9.5}$M$_\odot$, and the low-$z$~comparison from GAMA. We often use the rest-frame 0.5$\mu$m and 1.1$\mu$m S\'ersic index, $n_{0.5\mu m}$ and $n_{1.1\mu m}$, to examine the differences between rest-frame optical and rest-frame near-IR trends, choosing 0.5$\mu$m to avoid the larger scatter at shorter wavelength, and choosing 1.1$\mu$m to have common wavelength coverage across the entire redshift range.
    
\subsection{Wavelength dependence of S\'ersic n\label{sec: waveDep}}
    
    We divide the sample into quiescent and star-forming galaxies defining the former as those galaxies that are located $0.8$dex below the $SFR-M_*$ ridge defined by \cite{leja22} (see also Appendix \ref{Appendix}). The ridge definition presented in \cite{leja22} applies to $0.2<z<3$ galaxies. The SFRs of GAMA galaxies at $z<0.2$ are compared with the cutoff value for $z=0.2$ to avoid extrapolation. 

    Figure \ref{fig:sersicVSrestSF} shows the median S\'ersic index $n$~as a function of rest-frame wavelength.
    The medians (and percentile ranges) are constructed as follows. At a specific rest-frame wavelength all galaxies (minimum 10) with coverage from either HST or JWST are included using the Chebyshev polynomial value (see Sec.~2.3). If both HST and JWST cover that wavelength, JWST is preferred on account of its higher S/N.
    Lines are drawn using the \textsc{cobs} \citep{ng07,ng22} library which allows for a combination of a spline regression and quantile regression (the smoothing factor is chosen using the Schwarz-type information criterion automatically computed by the code.).
    
    We recover a clear, overall difference in the median S\'ersic index of star-forming and quiescent galaxies: 
    $1.36\pm0.01$~vs.~$3.28\pm0.03$, 
    respectively (as indicated by the grey dashed lines in Fig. \ref{fig:sersicVSrestSF}). Uncertainties are computed as standard deviation of 1000 replica of the medians computed at different wavelength and gaussian distributed according to their own statistical uncertainty (represented by the errorbars in Figure \ref{fig:sersicVSrestSF}).
    Neither the quiescent nor the star-forming population shows a strong wavelength dependence across $0.3\mu$m~to $2\mu$m. 
    The scatter in $n$ among all the galaxies, shown in the figure by means of shaded areas that extend from the $16^{th}$ to the $84^{th}$ percentiles, is constant with wavelength for quiescent galaxies, but for star-forming galaxies the scatter decreases with wavelength, likely due to a reduced variation in radial profiles due to stochasticity in the distribution of young stars and dust. This reduced scatter at long wavelengths implies that the contrast between quiescent and star-forming galaxies is, statistically speaking, more significant in the near-IR than in the UV/optical. 

    \begin{figure}[!t]
        \epsscale{1.15}
        \plotone{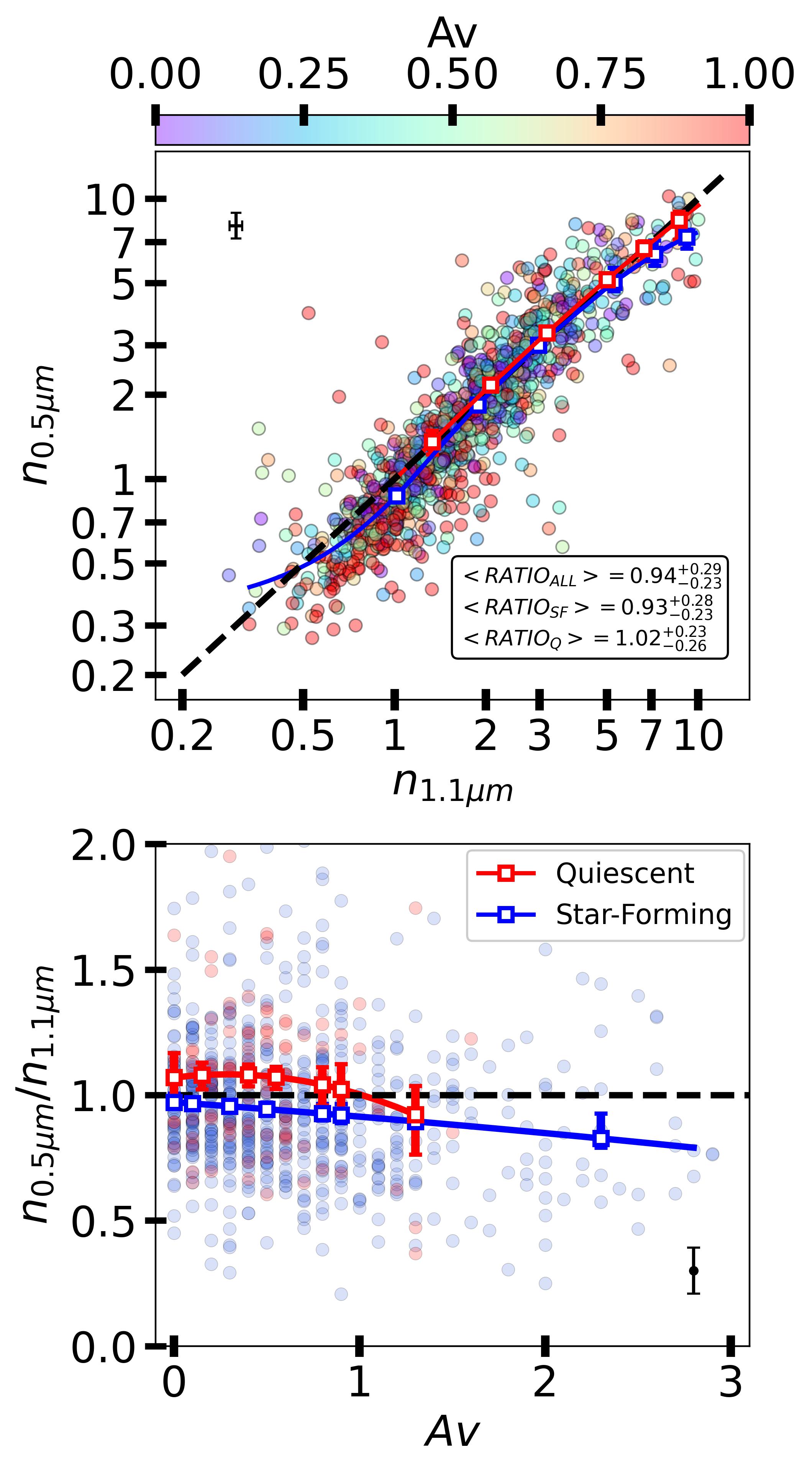}
        \caption{\emph{Upper panel:} S\'ersic index at $0.5\mu m$ against S\'ersic index at $1.1\mu m$ color-coded with $A_V$ (from the  \textsc{Prospector} SED fits). Blue and red solid lines show the median for star-forming and quiescent galaxies respectively computed as explained in the text. Error bars show the statistical uncertainty (84-16percentiles$/\sqrt{N}$). In the lower right corner is shown the median (and 16-84percentiles) ratio $n_{0.5\mu m}/n_{1.1\mu m}$ for the whole sample, the star-forming population and the quiescent population. In the upper left corner in black are shown the median uncertainties. \emph{Lower panel:} Ratio between S\'ersic index at $0.5\mu m$ and $1.1\mu m$ against $A_V$. Only galaxies with $SNR>100$ are included in this figure to highlight the trend with $A_V$. In the lower right corner in black is shown the median uncertainty propagated to the ratio of S\'ersic indices.
        \label{fig:n0.5_vs_n1.1_CC_AV}
        }
    \end{figure}

    The fact that the wavelength dependence is weak for the ensemble of galaxies does not necessarily imply that individual galaxies have similar S\'ersic indices at all wavelengths. In Figure \ref{fig:n0.5_vs_n1.1_CC_AV} we investigate the direct comparison between $n_{1.1\mu m}$~and $n_{0.5\mu m}$. The comparison shows only 0.1dex (26$\%$) scatter, which is a combination of observational uncertainties and physical variations. $19\%$ of galaxies change their S\'ersic index by more than a factor 1.5 (up or down) from $0.5\mu$m to $1.1\mu$m. The vast majority of galaxies have similar radial light profiles at different wavelengths. The fractions of galaxies with $n_{1.1\mu m}/n_{0.5\mu m}>1.5$ and $n_{0.5\mu m}/n_{1.1\mu m}>1.5$~are similar: 8\% and 11\%, respectively. At low $n$~there is a small but significant deviation in the sense that $n_{0.5\mu m}$ is smaller than $n_{1.1\mu m}$, driven by galaxies with large $A_V$. 
    We performed a Cram\'er-test \citep{baringhaus04} to asses whether $n_{0.5\mu m}$ and $n_{1.1\mu m}$ share the same distribution. For quiescent galaxies the test results 0.34 with an estimated p-value of 0.89 (with 1-$\sigma$ confidence interval) confirming the absence of wavelength dependence on S\'ersic index for quiescent galaxies. On the contrary, for star-forming galaxies, we find an observed statistic of 3.75 with a p-value of $10^{-3}$, suggesting the S\'ersic index indeed shows a mild dependence on wavelength. However, as shown in the bottom panel of Figure \ref{fig:n0.5_vs_n1.1_CC_AV}, $n_{0.5\mu m}/n_{1.1\mu m}$ becomes systematically smaller than unity for rising $A_V$. To statistically confirm this observation we performed a Cram\'er-test on those star-forming galaxies with Av$<0.3$ finding that $n_{0.5\mu m}$ and $n_{1.1\mu m}$ for these galaxies are distributed the same (test result 0.64, p-value 0.36).
    This is consistent with the recent results from \citet{gillman23} who find that sub-mm-selected galaxies (expected to be dusty) have more concentrated profiles in the rest-frame near-IR than in the rest-frame optical/UV.

     Figure \ref{fig:sersicVSrestSF} shows also the wavelength dependence of the S\'ersic index for the $z\leq0.3$~comparison sample drawn from GAMA (Sec.~\ref{sec:gama}). 
    Although the star-forming galaxies show an increase with wavelength from $n=1$ to almost $n=2$ in the rest-frame $H$ band, or $\approx$$45\%$ higher than seen in the high-$z$ sample, the quiescent population shows a milder evolution with its peak in the nUV.
    The trend shown by the star-forming galaxies echos the findings by \citet{Kelvin12} for GAMA, but they divided the sample into Disk and Spheroidal classes, presenting a different look compared to our separation by star-formation activity. It is worth pointing out that the definition of quiescence can be important: different definitions can affect the strength of the increase(decrease) in the nIR(nUV) shown by star-forming(quiescent) galaxies. Definitions like that presented in \cite{tacchella22} (where a galaxy starts its quiescent phase when sSFR$<$$1/(3t_H(z))$, with $t_H(z)$ being the age of the universe at the galaxy's redshift) lead to much shallower slopes for both the GAMA quiescent and star-forming population leaving almost unaffected our high redshift sample.
    
    Nonetheless, the good agreement between the two datasets suggests a lack of dependence on redshift for the quiescent population and a mild evolution in the median $n$ for the star-forming population.

    The trends shown in Figure \ref{fig:sersicVSrestSF} are for the full galaxy sample, with a wide range in redshift and stellar mass. Correlations with star-formation activity, stellar mass and redshift will be examined in further detail in Section \ref{sec: ssfrDep} and \ref{sec: massDep}.

\subsection{Correlations with Star Formation Activity \label{sec: ssfrDep}}
    
    \begin{figure*}[!t]
        \epsscale{1.05}
        \plotone{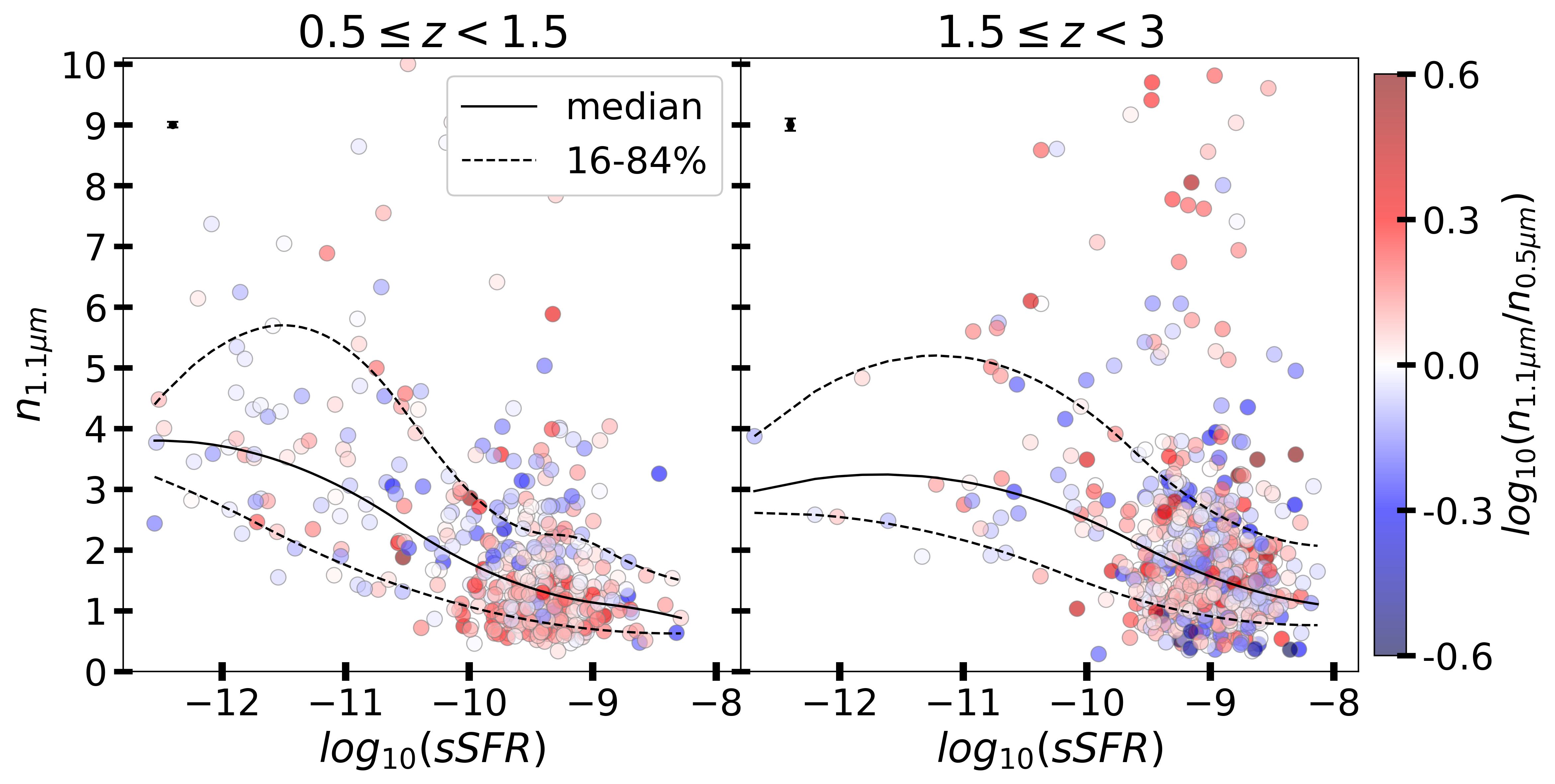}
        \caption{S\'ersic index at $1.1\mu m$ rest-frame against the sSFR in two redshift bins, color-coded with the ratio between the S\'ersic indices at $1.1\mu m$ and at $0.5\mu m$. The solid line shows the median for the whole sample in sSFR bins while the dashed lines show the 16-84 percentiles. In the upper left corner of each panel a black errorbar shows the median uncertainty of $n_{1.1\mu m}$.
        Low sSFR galaxies have systematically larger S\'ersic indices in the rest-frame near-IR.
        \label{fig:sersicVSssfr}
        }
    \end{figure*}

    Figure \ref{fig:sersicVSrestSF} suggests that star formation activity shows a strong correlation with S\'ersic index, regardless of wavelength.
    In Figure \ref{fig:sersicVSssfr} we show $n_{1.1\mu\rm{m}}$, the S\'ersic index at rest-frame  $1.1\mu$m, as a function of the sSFR.
    Lines are drawn with the technique presented in section \ref{sec: waveDep}. The smoothing factor is chosen using the Schwarz-type information criterion.
    Galaxies with low sSFR have systematically larger $n_{1.1\mu m}$ than galaxies with high sSFR. For both $n_{1.1\mu\rm{m}}$ and $n_{0.5\mu\rm{m}}$ the Cram\'er test excludes with very high confidence $<5\sigma$ that star-forming and quiescent galaxies are drawn from the same $n$ distribution.
    However, star-forming galaxies show a tail of high-$n$ galaxies. This echoes earlier results obtained in the rest-frame optical by \cite{bell12} and \cite{whitaker17}, who argued that having a high S\'ersic index is a necessary, but not sufficient, condition for a galaxy to be quiescent. 

    Interestingly, the S\'ersic index does evolve with redshift at fixed sSFR: galaxies with log(sSFR)$\sim -10$ have $n_{1.1\mu m}\lesssim 2$ at $z\sim 1$ and $n_{1.1\mu m}\sim 3$ at $z\sim 2$. But rather than indicating a physical decrease in $n_{1.1\mu m}$ with cosmic time, this should be interpreted in the context of a decline in SFR with cosmic time, which is of course well documented \cite[][and references therein]{leja22}.

    We also see that galaxies with low sSFR do not show a difference between $n_{1.1\mu m}$ and $n_{0.5\mu m}$.
    On the contrary, at $z<1.5$ (left-hand panel of Fig.~\ref{fig:sersicVSssfr}) high sSFR galaxies show positive values of $n_{1.1\mu m} / n_{0.5\mu m}$.
    At $z>1.5$ there is no such trend, but there is a significant scatter in $n_{1.1\mu m} / n_{0.5\mu m}$, which is mostly due to variations in $n_{0.5\mu m}$ (rather than $n_{1.1\mu m}$), otherwise this would result in a vertical gradient in the color coding. This scatter does not reflect difficulties in estimating S\'ersic profiles at $0.5\mu m$ at high redshift since these come from high-$S/N$~JWST/NIRCam imaging, but rather a wider variety of $n_{0.5\mu\rm{m}}$~values at high redshift, likely due to variance induced by bright star-forming regions and/or dust-obscured areas.

\subsection{Correlations with Stellar Mass \label{sec: massDep}}

    \begin{figure*}[!t]
        \epsscale{0.92}
        \plotone{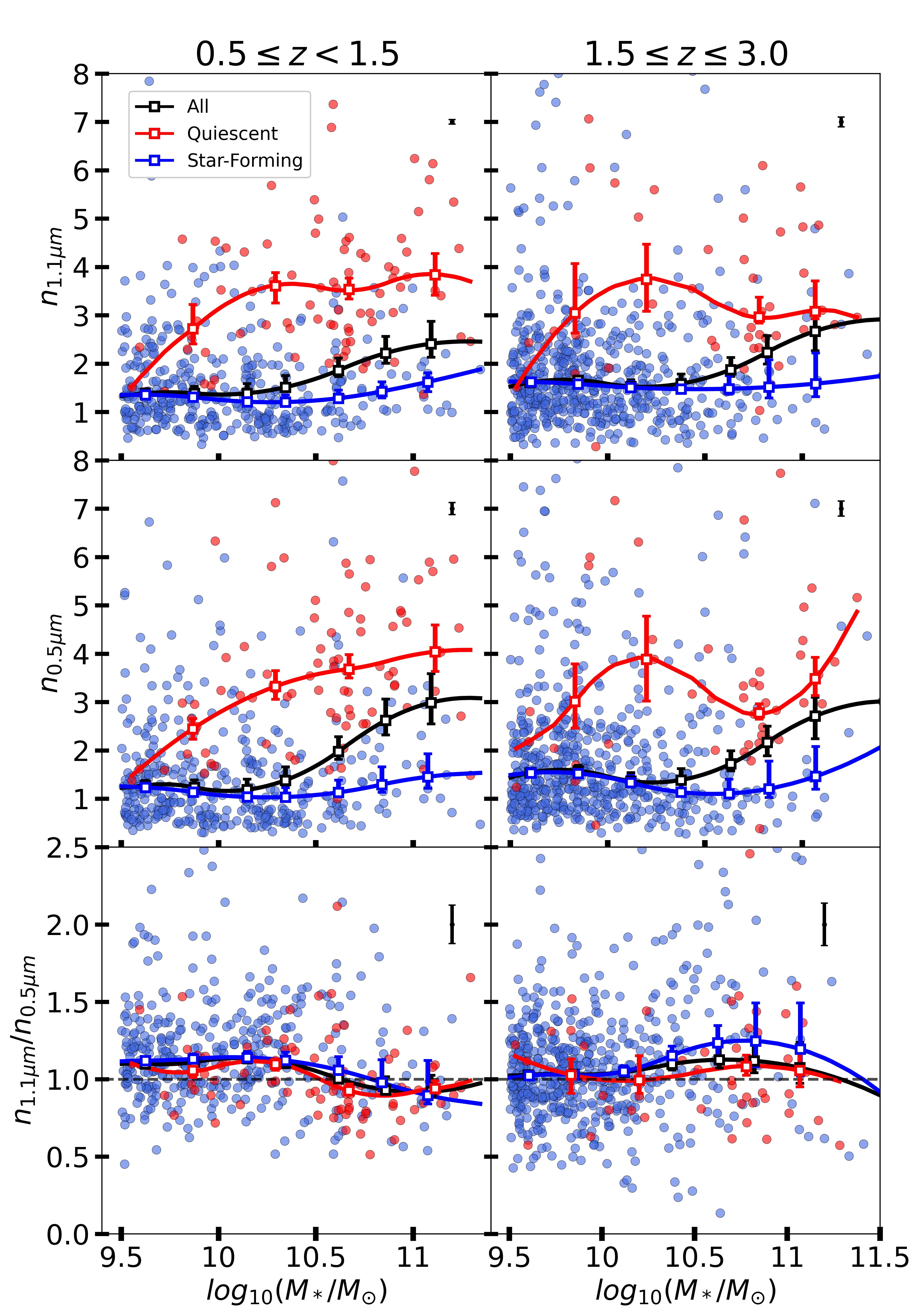}
        \caption{S\'ersic indices at $1.1\mu m$ (top row), $0.5\mu m$ (mid row) and their ratio (bottom row) vs. stellar mass, color-coded as star-forming (blue) and quiescent (red). The left column shows galaxies at $0.5\leq z<1.5$; the right column shows galaxies at $1.5\leq z\leq3$. Blue and red lines show the medians for the star-forming and quiescent galaxies,  respectively, while the black lines represent the full (combined) sample. Median lines are computed as described in section \ref{sec: waveDep} with a smoothing factor of 0.3 for the quiescent population and 1 for the star-forming and the whole sample. Errorbars show the statistical uncertainty (84-16percentiles$/\sqrt{N}$) in bins of width 0.5dex for the quiescent population and 0.25dex for the others. In the lower right corner of each panel a black errorbar represents the median uncertainty for the whole sample. We see a clear dependence of $n$~on stellar mass, regardless of wavelength and redshift, and no significant evidence for a mass- or redshift dependence for $n_{1.1\mu m} / n_{0.5\mu m}$.
        \label{fig:sersicRatioVSmass}
        }
    \end{figure*}
    
    The wavelength and sSFR dependence of the S\'ersic index shown above do not consider any correlation with stellar mass. 
    Figure \ref{fig:sersicRatioVSmass} shows $n_{1.1\mu m}$, $n_{0.5\mu m}$ and their ratio as a function of stellar mass. Regardless of wavelength, the S\'ersic index increase with stellar mass is in part due to the underlying correlations between, on the one hand, stellar mass and sSFR, and on the other hand sSFR and S\'ersic index. But also among quiescent galaxies the S\'ersic index increases with stellar mass, while for star-forming galaxies such an increase is less evident. Here we should keep in mind that the current sample has only a few massive star-forming galaxies with significant bulge components.

    Regardless of the detailed mass-dependencies that may exist, the main point is that the S\'ersic index - stellar mass distribution looks very similar in the rest-frame optical and rest-frame near-IR, and that these patterns exist across the entire redshift range $0.5<z<3$. This is further elucidated by the bottom panels, that show $n_{1.1\mu m}/n_{0.5\mu m}$, which does not deviate much from unity across the sampled stellar mass range. The only significant departure is seen, somewhat surprisingly, for lower-$M_*$ star-forming galaxies at $z<1.5$, which have slightly higher $n_{1.1\mu m}$ than $n_{0.5\mu m}$. These galaxies often have $n_{0.5\mu m}<1$ whereas $n_{1.1\mu m}\approx 1$, which may suggest that these galaxies have diffusely distributed young populations and/or somewhat attenuated centers, while the underlying populations have approximately exponential profiles.

    In the highest redshift bin we observe the median $n_{1.1\mu m}/n_{0.5\mu m}$ is compatible with unity for star-forming galaxies with $M_*<10^{10.3}$M$_{\odot}$ followed by an increase that peaks at $M_*\sim10^{10.7}$M$_{\odot}$ and a sharp decline for higher masses. The quiescent population shows no significant trend. However, the sample is too small at these high masses to claim a physical difference in $n_{1.1\mu m}/n_{0.5\mu m}$.

\section{Discussion \label{sec:  Conclusions-Discussion}}

    While light profiles in the optical/UV can be strongly affected by young stars and dust absorption, the near-IR is more representative of the stellar mass distribution. But the striking absence of a strong wavelength dependence of the S\'ersic index $n$~for both quiescent and star-forming galaxies over the redshift range $0<z<3$ suggests that the curvature of the radial light profile in the optical does not deviate much from that of the underlying stellar mass distribution across most of cosmic time: an exponential profile at short wavelengths predicts an exponential profile at long wavelengths, and a De Vaucouleurs-like profile generally also persists across the UV-to-near-IR wavelength regime.  Our findings reinforce the results from \citet{bell12} who found no wavelength dependence for $n$ across the rest-frame UV and optical wavelength regime for galaxies up to $z\sim 2.5$.
    
    The absence of a striking wavelength dependence on S\'ersic index  does not imply that light profiles do not systematically change with wavelength. 
    Galaxies have smaller scale radii at longer wavelengths at both low and high redshift \citep{Kelvin12, van-der-wel14}. At low redshift we understand this to be due to gradients in attenuation \citep[e.g.,][]{popescu2000, graham08_inclination} and stellar population properties \citep[e.g.,][]{sanchez-blazquez07, zibetti2020}. At large look-back time, thanks to JWST, such analysis is now becoming possible \citep{Miller2022, suess22, shen23}, and it is already clear that the stellar mass distribution is smoother \citep{wuyts12} than the clumpier distribution seen in the UV and optical \citep[e.g.,][]{guo15}. 

    The key result presented here is that the radial profiles of star-forming and quiescent galaxies are different even in the rest-frame near-infrared
    If star formation and/or dust were responsible for the different structure seen at shorter wavelengths, then the S\'ersic indices of star-forming and quiescent galaxies would become more similar at longer wavelengths. In fact, the reduced scatter in $n$ of star-forming galaxies at longer wavelengths ($\lambda_{\rm{rest}} \approx 1.5\mu$m, see Figure \ref{fig:sersicVSrestSF}) implies that the difference in structure is \emph{more} pronounced in the rest-frame NIR than the rest-frame UV/optical.
    The absence of such a trend implies, at first sight, that stellar mass profiles are similar to stellar light profiles, but since galaxy sizes decrease with wavelength, and stellar half-mass radii are generally found to be smaller than half-light radii \citep[e.g.,][]{szomoru13, fang13, suess17, mosleh17, suess19, miller23}, the interpretation is not straightforward. Clearly, the outshining effect of young, bright, blue populations \citep{wuyts12, reddy12, lilly16} plays a significant role. In particular, \citet{fang13} find that, for galaxies in the present-day Universe, the difference between half-mass and half-light radii is larger for star-forming galaxies than for quiescent galaxies and that stellar mass profiles in the range $3<R<10$~kpc are rather similar for quiescent and star-forming galaxies.
    
    But a high S\'ersic index is driven by the combination of deviations from an exponential profile at both small radius, where \citet{fang13} indeed find a significant difference between the mass profiles of star-forming and quiescent galaxies, and large radius, which \citet{fang13} do not examine. 
    In general, a decline in galaxy size with wavelength and a lack of such dependence for the S\'ersic index are not necessarily in tension, 
    and we conclude that there is a physical difference in the radial curvature of the stellar mass profiles when comparing star-forming and quiescent galaxies.
     
    The strong correlation between galaxy structure and star-formation activity seen at high redshift \citep{franx08, wuyts11,bell12,barro17, whitaker17} is now verified to be physical in nature and builds on well-documented correlations seen for present-day galaxies \citep[e.g.,][]{kauffmann03, brinchmann04}. A more in-depth discussion of the causal connection between structural evolution and star-formation history/quenching  \citep[e.g.,][]{van-der-wel09, bell12, fang13, van-dokkum15, lilly16, tacchella16, barro17, bluck20environmental, chen20, dimauro22} is beyond the scope of this paper.
    
    At $z\lesssim1.5$, $M_*<10^{10.3}$M$_{\odot}$ star-forming galaxies do show a \textit{mild} increase in $n$ with wavelength, while this trend disappears at higher redshift. Because of the lack of a significant number of high-mass galaxies in the sample, we can not argue the same for higher masses. However, the hint of a gradual trend of $n$ with redshift (Fig. \ref{fig:sersicRatioVSmass}) at first sight echoes the usual line of thought that older bulges and a lack of attenuation in the center lead to an increased S\'ersic index at longer wavelengths, but our measurements are not consistent with this picture. Instead, the trend is driven by galaxies with $M_*\lesssim10^{10}$M$_\odot$ which display a slight \textit{decrease} in $n$ from high to low redshift in the rest-frame optical (mid row of Fig.\ref{fig:sersicRatioVSmass}) while their rest-frame near-IR $n$ shows a weaker and less significant evolution (top row of Fig.\ref{fig:sersicRatioVSmass}). Similar trends were found at low redshift by \citet{vulcani14} and at $z\sim$2-3 by \citet{shibuya15} comparing UV and optical wavelengths. We speculate this is, at least partially, due to an increase in attenuation with cosmic time for galaxies in this mass range, associated with an increase in gas-phase metallicity \citep[e.g.,][]{sanders21}.

\section{Conclusions and Outlook \label{sec: Conclusions}}       

    We present rest-frame optical and near-IR S\'ersic index $n$ measurements for a sample of 1067 galaxies at $0.5<z<3$ with stellar masses $M_*$$\geq$10$^{9.5}$M$_\odot$ selected from recent JWST/NIRCam imaging that was collected as part of the CEERS program \citep{Finkelstein23}. The wavelength dependence of $n$ is  weak (Fig.\ref{fig:sersicVSrestSF}). As a result, the near-IR light profiles of galaxies do not, on average, strongly differ from those in the optical (Fig.\ref{fig:n0.5_vs_n1.1_CC_AV}). The large spread in $n$ at any given wavelength is strongly correlated with star-formation activity (Fig.\ref{fig:sersicVSssfr}), across redshift and stellar mass. Indeed, after controlling for star-formation activity, star-forming galaxies show just a weak evolution in $n$ with redshift, as previously shown by \citet{shibuya15} in the rest-frame optical, and a mild dependence on stellar mass. The remaining scatter is mostly driven by variations in the optical profile driven by dust absorption (Fig.\ref{fig:n0.5_vs_n1.1_CC_AV}). Regardless of wavelength, stellar mass, and redshift we see that star-forming galaxies have $n\sim1-1.5$, and quiescent galaxies have $n\sim 3.5$ (Figs.~\ref{fig:sersicVSrestSF} and \ref{fig:sersicRatioVSmass}).

    The fact that the well-documented correlation between galaxy structure and star-formation activity (see Sec.~\ref{sec: ssfrDep}) persists in the rest-frame near-infrared implies a physical connection between radial stellar mass distribution and growth through star-formation. In other words, it is not a mere perception caused by young, bright, star-forming disks that fade after the cessation of star formation.

    The lack of a strong wavelength dependence in the S\'ersic index at all redshifts $z<3$ also implies a plausible lack of such a dependence at even higher redshifts. As JWST/NIRCam explores the rest-frame optical up to $z\sim 10$ \citep{Kartaltepe22} and UV at $z>10$, we can have some confidence that the observed radial light profiles inform us about the underlying stellar mass profiles even if $M/L$-gradient corrections are needed to reconstruct those.
    
    Our conclusions hold for $0.1<L/L_*<1$ galaxies across the redshift range $0.5<z<3$ and the main weakness of the current study is the small number of high-mass star-forming galaxies (as discussed in section \ref{sec: massDep}), especially at high $z$: there are only 11(9) $M_*\geq10^{11}$M$_\odot$ star-forming(quiescent) galaxies at $1.5<z<3$ investigated with JWST in our sample. For these galaxies the structure may be expected to vary the strongest with wavelength due to centrally concentrated, dust-obscured star-formation activity. As larger areas are observed with NIRCam by, for example, JADES \citep{williams18}, UNCOVER \citep{bezanson22} and COSMOS-Web \citep{casey22}, this challenge will be addressed.

\section*{Acknowledgments}
    This project has received funding from the European Research Council (ERC) under the European Union’s Horizon 2020 research and innovation programme (grant agreement No. 683184). MM and MB acknowledge the financial support of the Flemish Fund for Scientific Research (FWO-Vlaanderen), research project G030319N.

\bibliographystyle{aasjournal}
\bibliography{mypapers}

\appendix 
\section{Sample Selection and Characteristics \label{Appendix}}
\begin{figure}[h!]
        \epsscale{1.15}
        \plotone{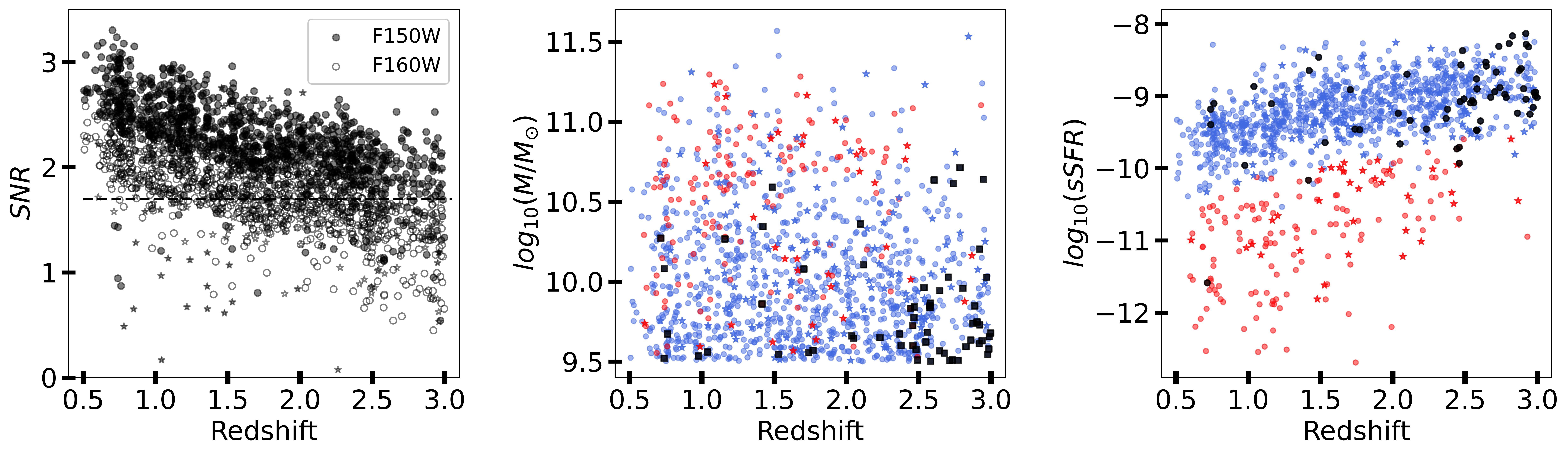}
        \caption{\emph{Left panel:} Signal-to-Noise Ratio (SNR) in the HST/ACS filter F160W (empty circles) and JWST/NIRCam F150W (filled circles) against redshift. The horizontal dashed black line shows the reference SNR=50. \emph{Central panel:} Stellar mass against redshift. \emph{Right panel:} specific Star Formation Rate against redshift.
        In the central and right panels colors red and blue are used to identify quiescenet and star-forming galaxies. Circles define galaxies used in this work while stars show galaxies rejected according to section \ref{sec: Data} and \ref{sec: Sersic Profile Fits}. 
        Black squares are galaxies with SNR$<50$ in F150W that are not rejected.
        \label{fig:Compl_and_outliers}
        }
    \end{figure}

\restartappendixnumbering
    Figure \ref{fig:Compl_and_outliers} shows the sample distribution and selection of galaxies used in this work. In the left panel we show the SNR in the HST/WFC3 filter F160W (empty circles) and JWST/NIRCam filter F150W (filled circles) as a function of redshift. The dashed horizontal line shows the SNR=50 limit suggested by \cite{van-der-wel12} to accurately measure S\'ersic indices.
    The central panel of Figure \ref{fig:Compl_and_outliers} shows the mass distribution of our galaxies across redshift. 
    The right panel of the figure shows the sSFR against the redshift distribution of our galaxies. In the last two panels star-forming galaxies are color-coded in blue while quiescent are red. The distinction between star-forming and quiescent is done as outlined in section \ref{sec: waveDep}. We used circles for galaxies included in the sample and stars for galaxies that are rejected as outlined in section \ref{sec: Data} and \ref{sec: Sersic Profile Fits}. Black squares represent galaxies that despite having a SNR$<50$ in F150W are not removed from the sample.
    
\renewcommand{\thefigure}{\arabic{figure}}

\end{document}